\documentclass[10pt]{article}
\usepackage{psfig}
\usepackage[cp1251]{inputenc}
\usepackage[russian]{babel}

\begin{document}

\twocolumn[ \noindent{\small\it Astronomy Reports, Vol. 49, No.
12, 2005, pp. 973--983. Translated from
Astronomicheski$\check{\imath}$ Zhurnal, Vol. 82, No. 12, 2005,
pp. 1087--1098. Original Russian Text Copyright \copyright 2005 by
Bajkova.}

\vskip -4mm

\begin{tabular}{llllllllllllllllllllllllllllllllllllllllllllllll}
 & & & & & & & & & & & & & & & & & & & & & & & & & & & & & & & & & & & & & & & \\
\hline \hline
\end{tabular}

\vskip 1.5cm

\centerline{\large\bf A Solution to the Problem of Phaseless
Mapping} \centerline{\large\bf for a High-Orbit Space-Ground Radio
Interferometer}

\bigskip

\centerline{\bf A. T. Bajkova}

\medskip

\centerline{\it Main (Pulkovo) Astronomical Observatory,
St.Petersburg, Russia}

\centerline{\small Received January 20, 2005; in final form, May
18, 2005}

\vskip 1.5cm

{\bf Abstract} --- {\small We consider the problem of mapping with
ultra-high angular resolution using a space--ground radio
interferometer with a space antenna in a high orbit, whose apogee
height exceeds the radius of the Earth by a factor of ten. In this
case, a multielement interferometer essentially degenerates into a
two-element interferometer. The degeneracy of the close-phase
relations prevents the use of standard methods for hybrid mapping
and self-calibration for the correct reconstruction of images. We
propose a new phaseless mapping method based on methods for the
reconstruction of images in the complete absence of phase
information, using only the amplitudes of the spatial-coherence
function of the source. In connection with this problem, we
propose a new method for the reliable solution of the phase
problem, based on optimizing information-carrying nonlinear
functionals, in particular,the Shannon entropy. Results of
simulations of mapping radio sources with various structures with
ultra-high angular resolution in the framework of the RADIOASTRON
mission are presented. \copyright {\it 2005 Pleiades
Publishing,Inc.} }

 \vskip 1.5cm
]

 \centerline{1.~INTRODUCTION }

\medskip

The problem of phaseless mapping in Very Long Baseline
Interferometry (VLBI) is considered in detail in [1], together
with the uniqueness of the solutions and methods for solving for
the phases. The current study is a logical continuation of this
work, since it is concerned with the phaseless mapping of radio
sources with ultra-high angular resolution using a ground--space
interferometer with a high-apogee orbiting antenna, and the
development of more trustworthy methods for the reconstruction of
images based only on the amplitudes of their Fourier spectra (the
spatial-coherence function).

\vskip 1.5mm

The problem of phaseless mapping in VLBI arises when the measured
phases are subject to large errors introduced by the medium
through which the radio wave propagates,and it is not possible to
use the closure phases and adaptive-calibration methods (hybrid
mapping and self-calibration [2]) traditionally applied in VLBI to
correctly reconstruct images.

\vskip 1.5mm

This is the case for a two-element interferometer [1], and also
for a ground--space radio interferometer with an antenna in a high
orbit whose apogee exceeds the Earth's radius by an order of
magnitude or more. In the latter case, the multi-element
interferometer essentially degenerates into two elements,
independent of the number of ground stations,leading to degeneracy
in the phases summed around triangles. As a result, mapping using
the standard methods becomes meaningless.

\vskip 1.5mm

This problem is relevant for the future Russian space mission
RADIOASTRON [3](planned to have an antenna in a high-apogee orbit
reaching heights of 350~000 km), intended to map extragalactic
radio sources with ultra-high angular resolution reaching
hundredths of a milliarcsecond (mas), with the goal of revealing
the workings of the central engines of distant quasars and
galaxies.

\vskip 1.5mm

As we indicated above,an effective strategy for phaseless VLBI
mapping was presented in [1]. This method is based on (1) making a
preliminary reconstruction of the amplitude of the visibility
function (spectrum) over the entire $UV$ plane by reconstructing
an intermediate image with zero spectral phase using the data
measured on a limited set of points, and (2) reconstructing the
desired image based on the amplitude of the source spectrum
obtained in the first stage using methods designed to solve the
phase problem.

\vskip 1.5mm

The goal of our current study is to develop a method for the
realization of the second step of this algorithm that is more
reliable than the method of Fienup that has been applied earlier.

\vskip 1.5mm

The goal of this work is topical. In spite of the fundamental
existence (apart from degenerate cases, defined on a set of
measure zero) of solutions of the phase problem for
multidimensional ($\ge 2$), spatially restricted signals described
by nonnegative real functions [4] that are unique with accuracy to
within a class of equivalent functions (a linear shift or rotation
of the image by 180$^o$), no trustworthy, practical algorithm has
been developed to obtain these solutions. For example, the
algorithms of Fienup [5,6], which are the most efficient and most
widely applied in practice, do not possess the property of
compression [7], so that they cannot guarantee convergence to the
correct solution in all cases. A successful application of Fienup
algorithms in the case of comparatively simple source structures
is demonstrated in [1].

\vskip 1.5mm

Here,we attempt to fill this gap by proposing a more fundamental
method for solving the phase problem, based on using the methods
of nonlinear optimization to search for global extrema.

\vskip 1.5mm

The following sections discuss VLBI mapping with an antenna in a
high orbit, describe the proposed method for solving the phase
problem, and present tests of the method and results of
simulations of the RADIOASTRON mission, aimed at mapping radio
sources with ultra-high angular resolution.

\bigskip

\centerline{2.~VLBI MAPPING WITH AN ANTENNA} \centerline{IN A HIGH
ORBIT}

\medskip

Placing at least one VLBI station beyond the Earth, and thereby
increasing the maximum baseline of the resulting interferometer,
makes it possible to appreciably enhance the resolving power of
the instrument.

\vskip 1.5mm

It is planned in the near future to realize the space project
RADIOASTRON [3] of the Russian Academy of Sciences (the Astro
Space Center, Lebedev Physical Institute), which aims to construct
a ground--space radio interferometer for the mapping of radio
sources with ultra-high angular resolution, of the order of
hundredths of a mas. Such resolutions are provided by observations
at 1.35 cm on the maximum baseline of 350 000 km, which is
achieved when the orbiting antenna is at apogee [8].

\vskip 1.5mm

The desire to obtain ultra-high angular resolution by increasing
the length of the baseline joining the ground stations and the one
orbiting station leads to certain mathematical problems in the
mapping, associated with the degeneracy of the multielement
ground--space interferometer into a two-element interferometer in
terms of the effective filling of the $UV$ plane, independent of
the number of ground stations [1]. One consequence of this is the
degeneracy of the relations for the closure phases that are
traditionally used in VLBI to correctly reconstruct the spectral
phases via adaptive-calibration methods.

\vskip 1.5mm

We can write the equations for the closure phases [2]:
$$
\widetilde C_{ijk}=\widetilde\phi_{ij}+\widetilde\phi_{jk}-
\widetilde\phi_{ik}=
$$
\begin{equation}
=\phi_{ij}+\phi_{jk}-\phi_{ik}+noise~term=C_{ijk}+noise~term,
\end{equation}
where
$\widetilde\phi_{ij}=\phi_{ij}+\theta_i-\theta_j+noise~term$,
$\phi_{ij}$ is the spectral phase of the source on the baseline
$(ij)$, $\theta_i$ and $\theta_j$ are the phases of the complex
gains of antennas $i$ and $j$, which include both instrumental and
atmospheric components that cancel out when the phases are summed
around a triangle, and "noise" term is the random, residual
component of the phase noise,which is usually small. Here, a tilda
denotes measured quantities.

\vskip 1.5mm

The degeneracy of the closure-phase equations is a consequence of
the geometrical degeneracy of the triangles, whose apices
correspond to the ground antennas $i$ and $j$ and the space
antenna $k$, which is very distant from the Earth. The right-hand
side of relation (1) essentially vanishes, independent of the real
spectral phase of the source: $\widetilde C_{ijk}\approx
noise~term \approx 0$. It is obvious that, in this situation,
applying the closure equations will always yield symmetrical
structures, independent of the real source structure.

\vskip 1.5mm

The problem of poor $UV$-plane coverage,which leads to large
sidelobes in the synthesized antenna beam, can be partially solved
by applying the technique of multifrequency synthesis [9,10].
However, this is not sufficient to correctly reconstruct the
phases. Therefore, we propose to use phaseless mapping based on
the presented methods to reconstruct the structure of the
source.We can correctly reconstruct the spatial orientation of the
source using a solution obtained via adaptive-calibration methods
based on observations with the ground-based (low-frequency) part
of the VLBI array.

\bigskip

\centerline{3.~METHOD FOR SOLVING}
\centerline{THE PHASE PROBLEM}

\medskip

Let us formulate the problem of reconstructing two-dimensional
images in discrete form. Let the discretization of the map be
carried out in accordance with the theorem of
Kotel'nikov--Shannon, and the dimensions of the map be $N\times
N$. The spectrum of the source is the $N$-point discrete Fourier
transform of the two-dimensional distribution $x_{ml}$ over the
source radiation with a finite carrier:

\begin{equation}
X_{nk}=\frac{1}{N}\sum_{m=0}^{N-1}\sum_{l=0}^{N-1} x_{ml}
\exp(\frac{-i 2\pi (nm+kl)}{N})=
\end{equation}

$$
 =A_{nk}+iB_{nk}=M_{nk}\exp(i\Phi_{nk}),
$$
where $A_{nk}$ is the real part, $B_{nk}$ the imaginary part,
$M_{nk}$ the magnitude, and $\Phi_{nk}$ the phase of the spectrum
$X_{nk}$, with

\begin{equation}
A_{nk}=M_{nk}\cos \Phi_{nk},~~~~~B_{nk}=M_{nk}\sin \Phi_{nk}.
\end{equation}

\vskip 1.5mm

We formulate this problem as follows. We wish to use known values
of the amplitude of a spectrum (the spatial-coherence function in
VLBI) $M_{nk}$ on some set of points in the spatial-frequency
domain (the $UV$ plane) to reconstruct the image $x_{ml}$, which
is equivalent to reconstructing the spectral phase $\Phi_{nk}$,
since the distribution $x_{ ml}$ is the inverse Fourier transform
of the total spectrum $X_{nk}$ (2), taking into account both the
amplitudes and phases.

\vskip 1.5mm

Let the measurements of the spatial-coherence function taking into
account (2) and (3) satisfy the relations

\begin{equation}
\sum_m\sum_l x_{ml}a_{ml}^{nk}+\eta_{nk}^r
=A_{nk}=M_{nk}\cos\Phi_{nk},
\end{equation}

\begin{equation}
\sum_m\sum_l
x_{ml}b_{ml}^{nk}+\eta_{nk}^i=B_{nk}=M_{nk}\sin\Phi_{nk},
\end{equation}
separately for the real and imaginary parts of the spectrum, where
$a_{ml}^{nk}=\cos(2\pi(mn+lk)/N)/N,~~b_{ml}^{nk}=\sin(2\pi(mn+lk)/N)/N$,
$\eta_{nk}^r$ and $\eta_{nk}^i$ are the measurement errors for the
real and imaginary parts of the spectrum, respectively, which obey
a Gaussian distribution with zero mean and dispersion
$\sigma_{nk}^2$.

\vskip 1.5mm

Let us represent the relations for the spectral phase as follows:

\begin{equation}
\cos\Phi_{nk}=2\cos^2(\Phi_{nk}/2)-1,
\end{equation}

\begin{equation}
\sin\Phi_{nk}=(\sin(\Phi_{nk}/2)+\cos(\Phi_{nk}/2))^2-1
\end{equation}
The reconstruction of the image can then be represented as the
solution of the following optimization problem with the linear
constraints (4), (5), into which we substitute the variables
$t_{nk}$ and $s_{nk}$ in accordance with (6)--(9):

\begin{equation}
\min~~ Q(x_{ml},t_{nk},s_{nk})+\sum_n\sum_k
\frac{({\eta_{nk}^r}^2+{\eta_{nk}^i}^2)} {2 \sigma_{nk}^2},
\end{equation}

\begin{equation}
\sum_m\sum_l x_{ml}a_{ml}^{nk}-M_{nk}t_{nk}+\eta_{nk}^r=-M_{nk},
\end{equation}

\begin{equation}
\sum_m\sum_l x_{ml}b_{ml}^{nk}-M_{nk}s_{nk}+\eta_{nk}^i=-M_{nk},
\end{equation}

\begin{equation}
 x_{ml}, t_{nk}, s_{nk}\ge 0,
\end{equation}
Here, $Q$ is a nonlinear functional determining the chosen
criteria for the quality of the reconstruction. The second term in
the optimization functional (10) is an estimate of the
disagreement between the solution and the measurements according
to an $\chi ^2$ criterion.

\vskip 1.5mm

In addition,it is easy to show that the variables $t_{nk}$ and
$s_{nk}$ satisfy the nonlinear constraint
\begin{equation}
\cos^2\Phi_{nk}+\sin^2\Phi_{nk}=(t_{nk}-1)^2+(s_{nk}-1)^2=1,
\end{equation}
which is key for the correct reconstruction of the spectral phase.

\vskip 1.5mm

Including the nonlinear constraints (14)in the Lagrange functional
in the standard way appreciably complicates the reconstruction
algorithm. We therefore propose the following scheme for the
solution of problems (10--(14). The problem with the linear
constraints (10)--(13) is first solved in the standard way via
direct optimization using Lagrange multipliers [7]. In the course
of the numerical iterative search for the extremum of the
corresponding dual functional (for example, using a
coordinate-descent method [11]), the nonlinear constraints (14)
are placed on the variables $t_{nk}$ and $s_{nk}$, similar to a
projection onto a convex set [7]. In this case, $t_{nk}$ and
$s_{nk}$ cannot vary independently of each other in their
determination of the spectral phase. This combined algorithm
incorporates the advantages of both methods for the optimization
of nonlinear functionals with linear constraints, characterized by
the presence of a global extremum, and iterative methods, which
are distinguished by the simplicity of their allowance for various
constraints on the solution directly in the computational
algorithm.

\vskip 1.5mm

When $Q$ is a functional of the Shannon entropy [12], namely,

\begin{equation}
Q(x_{ml},t_{nk},s_{nk})=\sum_m \sum_l x_{ml}\ln x_{ml}
\end{equation}
$$
+\sum_n\sum_k (t_{nk}\ln t_{nk}+s_{nk}\ln s_{nk})
$$
we have the maximum-entropy method (MEM). The basis for the
legitimacy of the functional (15), which is the total entropy of
the image and the new variables $t_{nk}$ and $s_{nk}$, can be
obtained based on a ray model for the formation of the image [7],
maximizing the joint probability for the formation of the image
and the field of the statistically independent variables
$\{t_{nk}, s_{nk}\}$.

\vskip 1.5mm

An important constraint appearing in the system of equations (11)
and specified by the necessary normalization of the image is the
constraint on the total flux of the source $M_o$ (the zeroth
harmonic of the spectrum):

$$
\sum_m\sum_l x_{ml}=M_{o}.
$$

\vskip 1.5mm

Another well-studied effective method is the minimum measure of
R\'{e}nyi [13]. In this case,the functional $Q$ has the appearance

$$
Q(x_{ml},t_{nk},s_{nk})=\sum_m \sum_l x_{ml}^{\alpha}+\sum_n
\sum_k (t_{nk}^{\alpha}+s_{nk}^{\alpha}),
$$
where $\alpha \ne 1$.

\vskip 1.5mm

In practice, in VLBI mapping,the Shannon entropy has become the
most widely accepted among the various nonlinear functionals
specifying various measures of the quality of reconstructed
images. Accordingly, we consider here a developed
phase-reconstruction method based on the MEM.

\vskip 1.5mm

Applying the standard Lagrange-multiplier method to problems
(10)--(13) and using (15) yields the following absolute
optimization problem:

\begin{equation}
\min~~L=\sum_m\sum_l x_{ml}\ln x_{ml}+\sum_n \sum_k (t_{nk}\ln
t_{nk}
\end{equation}

$$
+s_{nk}\ln s_{nk})+ \sum_n\sum_k
\frac{({\eta_{nk}^r}^2+{\eta_{nk}^i}^2)} {2 \sigma_{nk}^2}
$$

$$
+\sum_n \sum_k \alpha_{nk}(\sum_m\sum_l
x_{ml}a_{ml}^{nk}-M_{nk}t_{nk}
$$

$$
+\eta_{nk}^r+M_{nk})+\sum_n \sum_k \beta_{nk}(\sum_m\sum_l
x_{ml}b_{ml}^{nk}
$$

$$
-M_{nk}s_{nk}+\eta_{nk}^i+M_{nk}),
$$
where $\alpha_{nk}, \beta_{nk}$ are the Lagrange multipliers, or
the dual variables.

\vskip 1.5mm

The necessary condition for the existence of an extremum of the
functional $L$ is that the gradient $\frac{\partial{L}}
{\partial{y_{op}}} = 0$, and that the Hess matrix composed of the
elements $\frac{\partial^2{L}} {\partial{y_{op}}\partial{y_{qr}}}$
be positive semidefinite at the point where this gradient is zero
(here, the letter $y$ denotes the generalized variables
$x_{ml},t_{nk},s_{nk},\eta_{nk}^r,\eta_{nk}^i$, and $op,qr$ the
two-dimensional indices for these variables).

\vskip 1.5mm

A sufficient condition for the existence of a local extremum is
that the Hess matrix be positive definite. If the Hess matrix is
positive definite everywhere, the functional will be convex, and
the local extremum will be a global extremum.

\vskip 1.5mm

We can obtain a solution for the desired distribution (image)
$x_{ml}$ and the variables $t_{nk}$, $s_{nk}$ determining the
spectral phase from the necessary condition for the existence of
an extremum of the functional $L$:

\begin{equation}
x_{ml}=\exp(\sum_n\sum_k \alpha_{nk} a_{ml}^{nk}+\beta_{nk}
b_{ml}^{nk}-1),
\end{equation}

\begin{equation}
t_{nk}=\exp(\alpha_{nk}M_{nk}-1),~~~s_{nk}=\exp(\beta_{nk}M_{nk}-1),
\end{equation}

$$
\eta_{nk}^r=-{\alpha}_{nk}{\sigma}^2_{nk}~~~\eta_{nk}^i=-{\beta}_{nk}{\sigma}^2_{nk}.
$$
As we can see from (17)and (18), the solutions for the variables
$x_{ml},t_{nk},s_{nk}$ are always positive; i.e., the condition
(13) is satisfied automatically, which is an internal property of
the entropy functional. It is not difficult to show that the Hess
matrix is diagonal and has positive elements, which means that it
is positive definite everywhere, so that the Lagrange functional
(16) is convex and the solution is global, i.e., unique.

\vskip 1.5mm

However, the uniqueness of the solution in this case refers only
to the uniqueness of the reconstruction of the shape of the
source. Images obtained as  a result of linear shifts or $180^o$
rotations of the solution (17) also satisfy (10)--(14), as follows
from the properties of the trigonometric functions determining the
desired phase in terms of the variables $t_{nk}$ and $s_{nk}$ [see
(8),(9)], and so are likewise solutions of the functional (16).
All these solutions comprise a class of equivalent functions that
differ by linear shifts or rotations by $180^o$. Thus, the
proposed method for solving the phase problem yields a unique
solution with accuracy to within a class of equivalent functions.
We used the coordinate-descent method studied in detail in [11]
for various functionals to numerically realize the absolute
optimization (16).

\vskip 1.5mm

In addition, note that the proposed mapping method based on the
MEM possesses higher stabilizing properties with respect to the
noise than, for example, the CLEAN algorithm that is traditionally
used in adaptive-calibration methods in VLBI, or the method of
Fienup used in [1]. The high stability of the proposed method is
due to both the properties of the nonlinear entropy functional and
the possibility of including the real signal-to-noise ratio in the
$\chi^2$ criterion [see (10)][14].

\bigskip

\centerline{4.~ TESTING THE METHOD}

\medskip

We tested the proposed method using the eight model radio sources
shown in Fig.1. The models and the radio sources themselves will
be referred to in accordance with the notation in the figure,
using the letters "a--h". These models reflect the various
characters of possible brightness distributions --- from a
collection of unresolved point sources and various numbers of
Gaussian components with various relative positions, to uniform
distributions within specified boundaries and rings (see the
following section for more detail on these source structures).
Note that the lowest contour and the constant step between
contours in all images is 1\% of the peak of the map.

\vskip 1.5mm

Figure 2 shows the images reconstructed in the case of zero
spectral phase. As was noted in Section 2, this is characteristic
of situations in which the phase triangles are completely
degenerate. These images were produced using standard mapping
methods. We can see that the source structures have taken on a
symmetrical form. Mapping methods that enable  reconstruction of
the spectral phase are required if we wish to derive the intrinsic
structure of the source. Images obtained using our proposed method
to reconstruct the phases are presented in Fig.3.

\vskip 1.5mm

We took the full set of measurements of the amplitude of the
spectrum as the input data. A comparison of the model (Fig.1) and
reconstructed (Fig.3) maps demonstrates the fairly high internal
accuracy of the method, which is from 1\% to 4\% for these
examples. We can see from Fig.3 that some sources are
reconstructed with accuracy to within a linear shift ("g" and "h")
or within a rotation by $180^o$ ("c" and "d").

\vskip 1.5mm

To test the application of the proposed method to real VLBI data,
we carried out phaseless mapping of the radio source 2200 +420
using data obtained during a global astrometric/geodetic observing
program at 8.2 GHz (wavelength 3.5 cm) in 1996--1997. The
$UV$-plane coverages corresponding to these data are presented in
[1]. Figure 4 shows images reconstructed with angular resolutions
of the order of 0.5--0.7 mas. These maps qualitatively and
quantitatively agree with the results obtained in [1] using
alternative mapping methods --- both phaseless methods based on
the algorithm of Fienup and standard methods based on
self-calibration. Real geodetic VLBI data are not intended for
astrophysical mapping, and therefore do not have the highest
quality in terms of calibration and signal-to-noise ratio.
Nevertheless, our analysis of maps made from such data shows that
the proposed method is able to construct high-quality images with
resolutions determined by the geometry of the interferometer.

\bigskip

\centerline{5.~SIMULATIONS} \centerline{OF THE RADIOASTRON
MISSION}

\medskip

We will now present the results of simulating mapping radio
sources using a ground--space VLBI system with the following
parameters.

\vskip 1.5mm

The ground stations used (the choice is not of fundamental
importance here) were the Svetloe, Zelenchuk, and Badary stations
of the QUASAR network. The wavelength of the simulated
observations was 1.35 cm. In the case of multifrequency synthesis,
we used a frequency band with a width equal to 30\% of the
frequency corresponding to the chosen wavelength. The orbit of the
RADIOASTRON spacecraft [8] is inclined to the equatorial plane by
$51.5^o$ and has an angle from the Vernal Equinox to the line of
nodes of $-45^o$, an angle from the line of nodes to perigee of
$30^o$, and perigee and apogee heights of 20 000 and 350 000 km,
respectively. The period of the spacecraft in its orbit around the
Earth is 9.5 days.

\vskip 1.5mm

Figures 5a and 5b depict the coverage of the $UV$ plane obtained
for the cases of single-frequency and multifrequency syntheses
over a time equal to one period of the spacecraft around the
Earth. The spatial frequencies are plotted along the $U$ and $V$
axes in units of $10^8$ wavelengths.

\vskip 1.5mm

Note that, since the bandwidth for the multifrequency synthesis
images was taken to be no more than 30\% of the central frequency,
the effect of the frequency dependence on the images can be
neglected. According to the estimates of [15], the influence on
the synthesized image due to this effect is no more than 1\%, even
when the source has a large spectral index. Correcting for the
frequency can lower this influence to 0.1\%. Our aim here is only
to demonstrate the effect of the multifrequency synthesis regime
on the coverage of the $UV$ plane, compared to the case of
single-frequency synthesis. If a wider band is used for the
frequency synthesis, this requires application of frequency
corrections, in accordance with the well known algorithms
described in [15].

\vskip 1.5mm

These parameters for the mapping system correspond to a maximum
angular resolution of $\lambda/D =0.007$ mas, where $\lambda$ is
the wavelength and $D$ is the maximum baseline of the
ground--space interferometer. The resolution in the east--west
direction is roughly half that in the north--south direction. The
following simulations will show that, thanks to the application of
nonlinear methods, it is possible to reconstruction the images
with a resolution higher than that indicated above.

\vskip 1.5mm

As the model radio sources depicted in Fig.1, we used simplified
Gaussian models derived from maps on mas scales constructed from
VLBA [16] data at 8.2 GHz, but translated to the angular scales
considered here (tens of microarcsecond).

\vskip 1.5mm

Thus, the model-source structures presented in Fig.1 represent (a)
a collection of point features corresponding to the three
brightest components in the source 1223+395 [17], (b) a core and
short curved jet adjacent to the core (a model for the quasar
0215+015), (c) an extended core and a comparatively bright
one-sided, extended jet (a model for the quasar 1607+268), (d) a
core and weak, one-sided, extended jet (a model for the quasar
1022+194), (e) a core and two-sided, multicomponent, bright jet (a
model for the quasar 0238$-$084), (f) a collection of several
bright and weak Gaussian components in a region about 50
microarcsecond in diameter (a model for the unidentified radio
source 0259+121), (g) a uniform brightness distribution within
specified boundaries, and (h) a thin ring. These two last source
structures ("g" and "h") are the most exotic and demanding in
terms of obtaining acceptable images by extrapolating the spectral
data to the high-frequency domain, and are considered here in
order to demonstrate the limits of the proposed reconstruction
method.

\vskip 1.5mm

Figures 6 and 7 present the images reconstructed based on the
amplitude data for both the single-frequency and multifrequency
syntheses. Note that we added to the data uniform noise in the
range $\pm 10$\% of each measurement of the visibility function,
which yielded an input signal-to-noise ratio for the algorithm of
about ten, in agreement with the sensitivity parameters for the
RADIOASTRON mission presented in [18].

\vskip 1.5mm

We give here a qualitative analysis of the reconstruction results.
We will not take into consideration rotations of the images by
$180^o$. We will refer to single-frequency and multifrequency
syntheses as the first and second cases, respectively.

\vskip 1.5mm

{\it Source "a".} The quality of the reconstruction was slightly
worse in the first than in the second case. We can see several
false point features (artefacts), but their brightnesses are
appreciably lower than the brightnesses of the correctly
reconstructed features (<4\% of the peak value). More than the
other distortions, the false features that are located along a
line passing through the center of the source manifest the
incomplete reconstruction of the spectral phase. In the second
case, these artefacts are not present. In the first case, the
shape of the source and the amplitude ratios of the features are
somewhat distorted, while it was possible to reconstruct the
correct amplitude ratios in the second case. The brightnesses of
the small number of point artefacts in the second case is no more
than 1--2\%. The point features were reconstructed with finite
resolution, but with a resolution that is higher than that of the
array used. The resolution in the north--south direction is about
twice that in the east--west direction. In both cases, the
coordinates of the source components were accurately
reconstructed. This simulation is useful because it clearly
demonstrates the resolving capability of the reconstruction
method.

\vskip 1.5mm

{\it Source "b".} In both cases, it was possible to reconstruct
the core well, but the adjacent jet much more poorly; in the
second case, however, the jet is manifest more distinctly. In the
first case, the features are elongated in the upward direction (in
amplitude), making the weak, extended feature virtually invisible.
This can be explained by the redistribution of the high-frequency
components of the spectrum as a result of the extrapolation of the
data belonging to a limited number of tracks in the $UV$ plane
[1]. The larger the region occupied by the data, the more exact
the values of the extrapolated spectrum. The level of false
extended features in the maps does not exceed 2\%.

\vskip 1.5mm

{\it Source "c".} As in the previous example, the Gaussian
components are elongated in amplitude in the first case, likewise
due to the redistribution of the spectrum at high spatial
frequencies. The reconstruction of the shapes of the components
and their brightness ratios is more accurate in the second case.
The level of false features does not exceed 1\%.

\vskip 1.5mm

{\it Source "d".} A fundamental difference of this source
structure from that considered in the previous example is that the
brightness of the jet is appreciably lower than that of the core.
In this case, the reconstruction of the image is more complex,
since the spectral phases are expressed less strongly. This was
visible in the results of the simulations, which are presented in
the table (as an example of the most typical type of radio
source). Compared to the previous simulation, the incomplete
reconstruction of the phases led to a growth in the brightnesses
of the false components, which could be taken to be a counterjet.
In the second case, however, the quality of the image is much
better; the shapes and amplitude ratios of the components are
reconstructed  more accurately,and the level of artefacts is
reduced by a factor of two.

\vskip 1.5mm

{\it Source "e".} In the first case, the direction of the jet is
reconstructed well, but individual components in the jet are
expressed appreciably worse (due to the lower resolution in the
east--west direction) than in the model, although they are
nonetheless resolved. The maximum level of extended false features
is about 10\%. In the second case, both the shapes (although there
is some elongation) and amplitude ratios of the components are
reconstructed well. The artefacts occupy fewer regions, and their
level does not exceed 5\%.

\vskip 1.5mm

{\it Source "f".} In these simulations, we have approximately the
same relative qualities of the maps obtained for the
single-frequency and multifrequency synthesis data as in the
previous example. In the first case, there is some smearing near
adjacent components due to the lower resolution in the east--west
direction. All the components except for the weakest were
reconstructed fairly well in the second case, although we can also
see some elongation of their shapes. The weakest pointlike
component was not able to be reconstructed in either case,
reflecting the comparatively low dynamical range of the resulting
maps.

\vskip 1.5mm

{\it Source "g".} In both cases, we have a fairly good
reconstruction of the boundary of the source, although there are
three fairly bright artefacts (about 8\% of the peak brightness)
adjacent to the outer boundary in the first case. The resulting
brightness distribution inside the boundary is not uniform. The
brightness of artefacts is appreciably lower in the second case;
they do not press up against the boundaries of the source and they
have a pointlike character. The brightness distribution inside the
source is substantially more uniform.

\vskip 1.5mm

{\it Source "h".} The shape of the ring is reconstructed more
distinctly and with a larger number of features in the second
case. In both cases, there are pointlike artefacts adjacent to the
ring, but the shape of the ring is clearly traced.

\vskip 1.5mm

Thus, our analysis of the results of each simulation demonstrates
that we were able to obtain an acceptable reconstruction, even in
the case of poor $UV$-plane coverage (the single-frequency
synthesis), at least for those sources with comparatively simple
core + jet structures (such as sources "c", "d"). The
multifrequency synthesis provided appreciably more accurate
reconstructions than the single-frequency synthesis, with this
difference being more pronounced as the source structure became
more complex (see the models for sources "e"$-$"h"). In all eight
of the simulations, the multifrequency synthesis enabled more
accurate reconstruction of the shapes and amplitude ratios of the
components, provided higher resolution, and lowered the level of
artefacts.

\vskip 1.5mm

Analysis of the results (Figs.6 and 7) shows that the
reconstructed maps have a resolution exceeding that of the
instrument, especially in the case of multifrequency synthesis
(the maps of sources "e" and "f"), although a ratio of 2:1 is
preserved for the resolution in the north--south and east--west
directions. The achievement of higher resolution is due to the
fact that the image-reconstruction method is based on the
substantially nonlinear maximum-entropy method, which has the
property of super-resolution; i.e., the ability to trace the
spectrum outside the aperture of the instrument. This frequency
extrapolation is more accurate in the case of multifrequency
synthesis, due to the more complete coverage of the $UV$ plane.

\vskip 1.5mm

If we convolve the solutions obtained (Figs.6 and 7) with a
"clean" beam corresponding to half the width of the main lobe of
the synthesized "dirty" beam for the system, in order to use the
most reliably reconstructed part of the spectrum and to avoid the
effect of overdetermination of the spectrum, on average, we obtain
maps with angular resolution of the order of 0.01 mas.

\vskip 1.5mm

Thus,when measuring the visibility function with a signal-to-noise
ratio of about ten using a radio interferometer with the geometry
specified for the RADIOASTRON mission, we can achieve a resolution
of no worse than 0.010 mas at 1.35 cm. At lower signal-to-noise
ratios, an appreciable role in the optimized functional (10)
begins to be played by the regularizing term $\chi^2$, which can
lead to some loss of resolution. Estimates show that this loss
will constitute a factor of about 1.5 for a signal-to-noise ratio
of about three.

\vskip 1.5mm

It follows that, when used together with multifrequency synthesis,
the proposed method for phaseless mapping can be successfully
applied to obtain maps of good quality on angular scales of the
order of tens of mas for data obtained on the RADIOASTRON
ground--space interferometer, when standard methods of adaptive
calibration can not be used due to the degeneracy of the
closure-phase equations.

\bigskip

\centerline {6.~CONCLUSION }

\medskip

The task of phaseless VLBI mapping is especially topical now,in
connection with future radio interferometers that may include a
space antenna in a high orbit whose apogee exceeds the radius of
the Earth by a factor of several dozen or more. In this case, the
closure-phase relations become degenerate, making application of
the standard methods of adaptive calibration incorrect. Our
results on simulations for the RADIOASTRON ground--space radio
interferometer, designed for imaging with ultra-high angular
resolution, demonstrate the ability to achieve acceptable image
reconstructions when our proposed phaseless mapping methods are
used together with multifrequency synthesis techniques.

\bigskip

\centerline {ACKNOWLEDGMENTS}

\medskip

The author is grateful to N.S.Kardashev and A. V. Stepanov for
support of this work. This work was partially supported by the
Basic Research Program of the Presidium of the Russian Academy of
Sciences "Nonstationary Phenomena in Astronomy".

\bigskip

\centerline{REFERENCES}

\medskip
{\small

\noindent 1.~A.~T.~Baikova, Pis ’ma Astron.Zh. {\bf 30}, 253
(2004) [Astron. Lett. {\bf 30}, 218 (2004)].

\noindent 2.~T.~J.~Cornwell and E.~B.~Fomalont, in {\it Synthesis
Imaging in Radio Astronomy II. A Collection of Lectures from the
Sixth NRAO/NMIMT Synthesis Imaging Summer School}, Ed. by
G.~B.~Taylor, C.~L.~Carilli, and R.~A.~Perley (Astron. Soc. Pac.,
San Francisco, 1999); Astron. Soc. Pac. Conf. Ser. {\bf 180}, 187
(1999).

\noindent 3.~N.~S.~Kardashev, Exp. Astron. {\bf 7}, 329 (1997).

\noindent 4.~Yu.~M.~Bruck and L.~G.~Sodin, Opt. Commun. {\bf 30},
304 (1979).

\noindent 5.~J.~R.~Fienup, Opt. Lett. {\bf 3}, 27 (1978).

\noindent 6.~J.~R.Fienup,~Appl. Opt. {\bf 21}, 2758 (1982).

\noindent 7.~G.~I.~Vasilenko and A.~M.~Taratorin, {\it Image
Restoration} (Radio i Svyaz’, Moscow, 1986)[in Russian].

\noindent 8.~{\it RADIOASTRON (Space VLBI Project),}\\
http://~www.~asc.~rssi.~ru/radioastron/description/\\orbit\_eng.htm.

\noindent 9.~{\it RADIOASTRON (Space VLBI Project),}\\
http://www.asc.rssi.ru/radioastron/description/\\mfs\_eng.htm.

\noindent 10.~A.~T.~Bajkova, Preprint No. 24, IPA AN SSSR (Inst.
Appl. Astron., USSR Acad. Sci., Leningrad, 1990).

\noindent 11.~A.~T.~Bajkova, Preprint No. 58, IPA RAN (Inst. Appl.
Astron., Russian Acad. Sci., St.-Petersburg, 1993).

\noindent 12.~B.~R.~Frieden, J. Opt. Soc. Am. {\bf 62}, 511
(1972).

\noindent 13.~B.~R.~Frieden and A.~T.~Bajkova, Appl.Opt. {\bf 34},
4086 (1995).

\noindent 14.~T.~J.~Cornwell, R.~Braun, and D.~S.~Briggs, in {\it
Synthesis Imaging in Radio Astronomy II. A Collection of Lectures
from the Sixth NRAO/NMIMT Synthesis Imaging Summer School}, Ed. by
G.~B.~Taylor, C.~L.~Carilli, and R.~A.~Perley (Astron. Soc. Pac.,
San Francisco, 1999); Astron. Soc. Pac. Conf. Ser. {\bf 180}, 151
(1999).

\noindent 15.~R.~J.~Sault and J.~E.~Conway, in {\it Synthesis
Imaging in Radio Astronomy II. A Collection of Lectures from the
Sixth NRAO/NMIMT Synthesis Imaging Summer School}, Ed. by
G.~B.~Taylor, C.~L.~Carilli, and R.~A.~Perley (Astron. Soc. Pac.,
San Francisco, 1999); Astron. Soc. Pac. Conf. Ser. {\bf 180}, 419
(1999).

\noindent 16.~C.~Ma and M.~Feissel, IERS Technical Note No. 23
(1997).

\noindent 17.~D.~R.~Henstock, I.~W.~Browne, P.~N.~Wilkinson, {\it
et al.}, Astrophys. J., Suppl. Ser. {\bf 100}, 1 (1995).

\noindent 18.~{\it RADIOASTRON~(Space~VLBI~Project),}\\
http://www.asc.rssi.ru/radioastron/description/\\observations\_eng.htm.

}

\bigskip

{\it Translated by D.Gabuzda}

\clearpage \onecolumn

\begin{small}
\centerline{ Characteristics of the measurements and reconstructed
images}

\begin{center}
\noindent\begin{tabular}{|c|c|c|c|c|c|} \hline
              &Peak input noise relative &Signal-to-noise  &Peak relative&Entropy   &Peak false    \\
  Map         &to the measured amplitude &ratio            &brightness   &of the map&feature, \%   \\
              &of the visibility function&in the input data&of the map   &          &              \\
\hline
Fig.~1d        &       --       &    --      &  1.00       &  -28.65&    --         \\
\hline
Fig.~6d        &  0.1           &   11.67    &  0.93       &  -31.74&  6      \\
\hline
Fig.~7d        &  0.1           &   10.04    &  1.06       &  -33.07&  3      \\
\hline
\end{tabular}
\end{center}
\end{small}

\begin{figure}[d]
\centerline{\psfig{figure=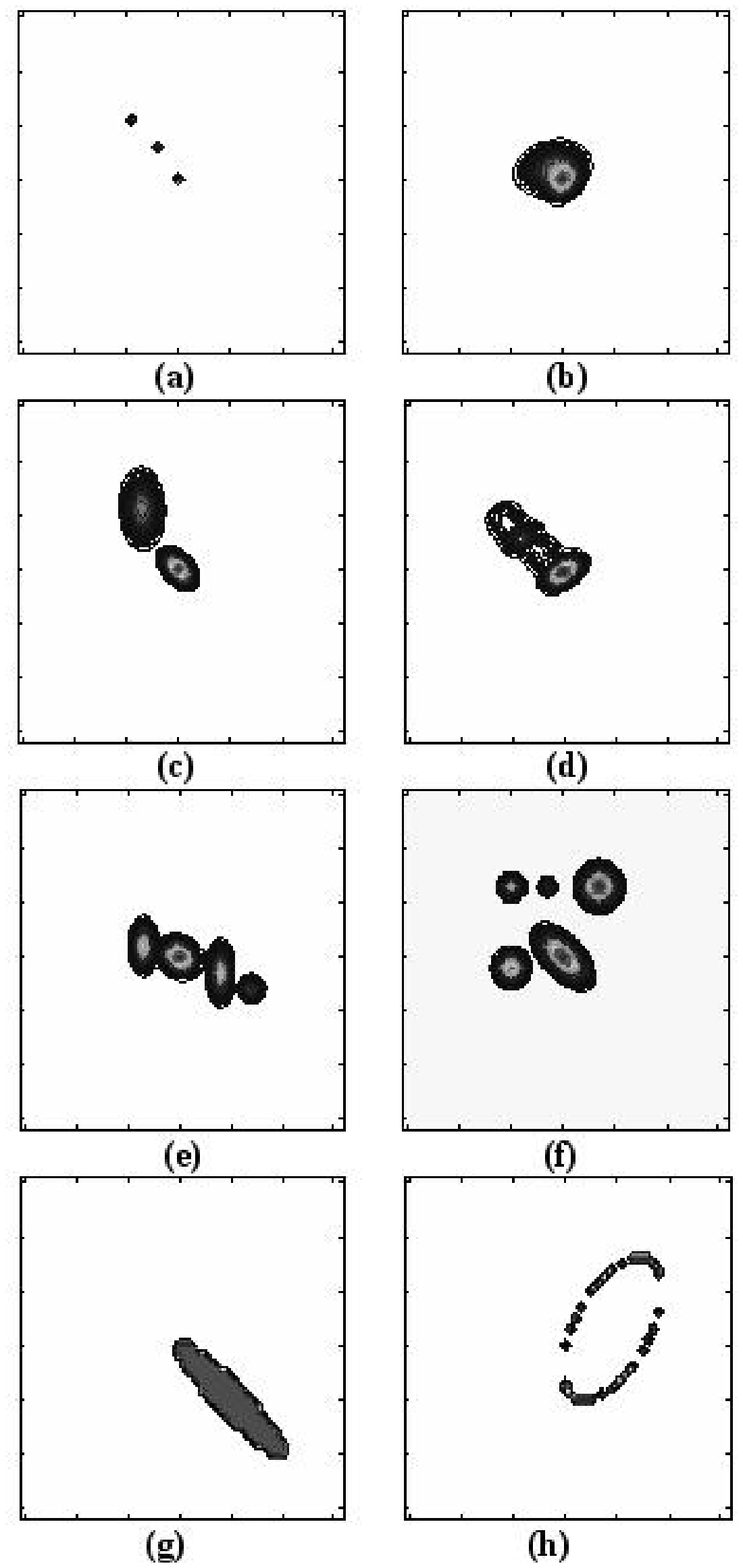,angle=0,width=100mm}}

\vskip 2mm

\begin{center}
{\bf Fig.1.}~Specified images of model radio sources.
\end{center}
\end{figure}

\begin{figure}[d]
\centerline{ \psfig{figure=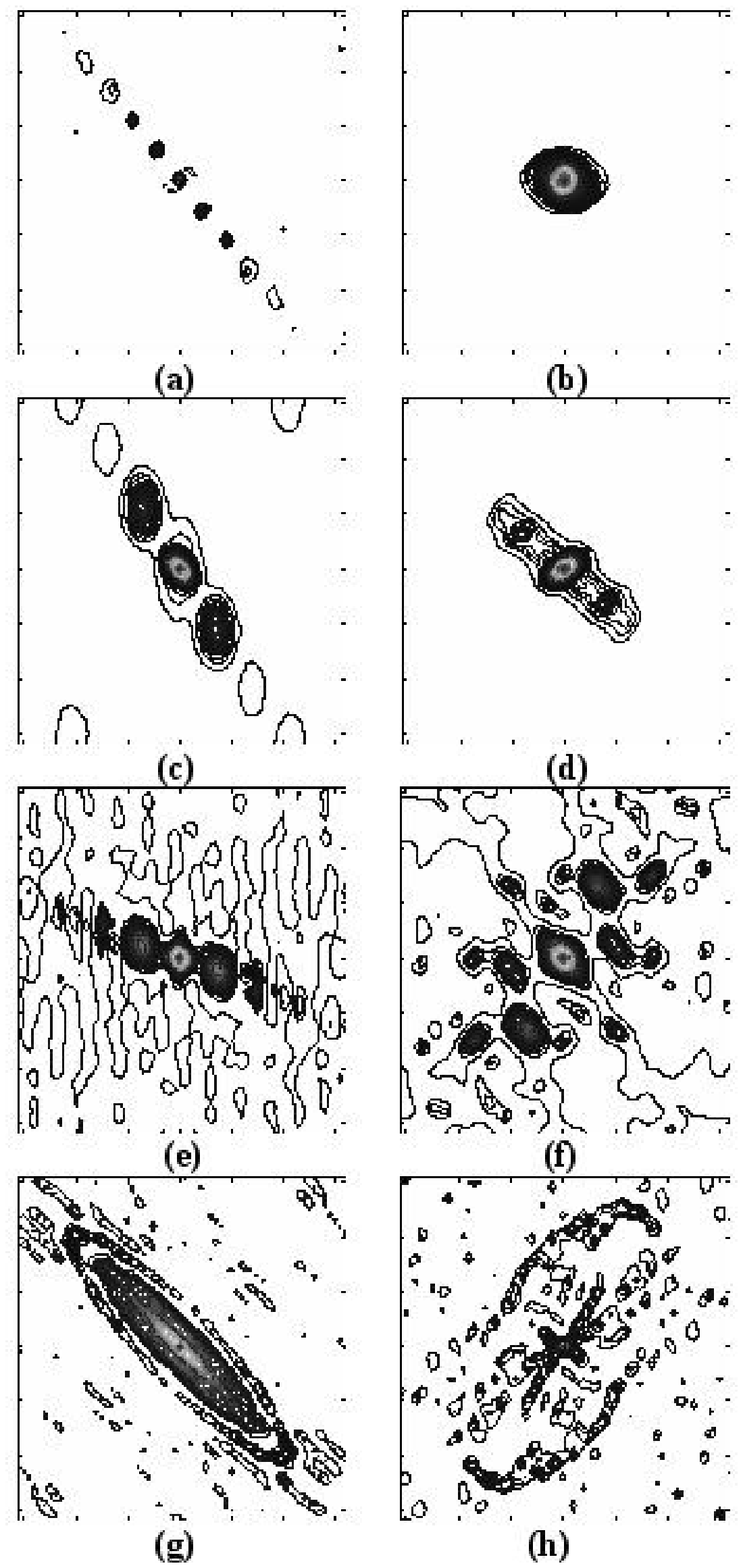,angle=0,width=100mm} }

\vskip 2mm

\begin{center}{\bf Fig.2.}~Images of model radio sources
reconstructed \\ using standard mapping methods in the case of\\
degenerate phase triangles (zero spectral phase).
\end{center}
\end{figure}

\begin{figure}[d]
\centerline{ \psfig{figure=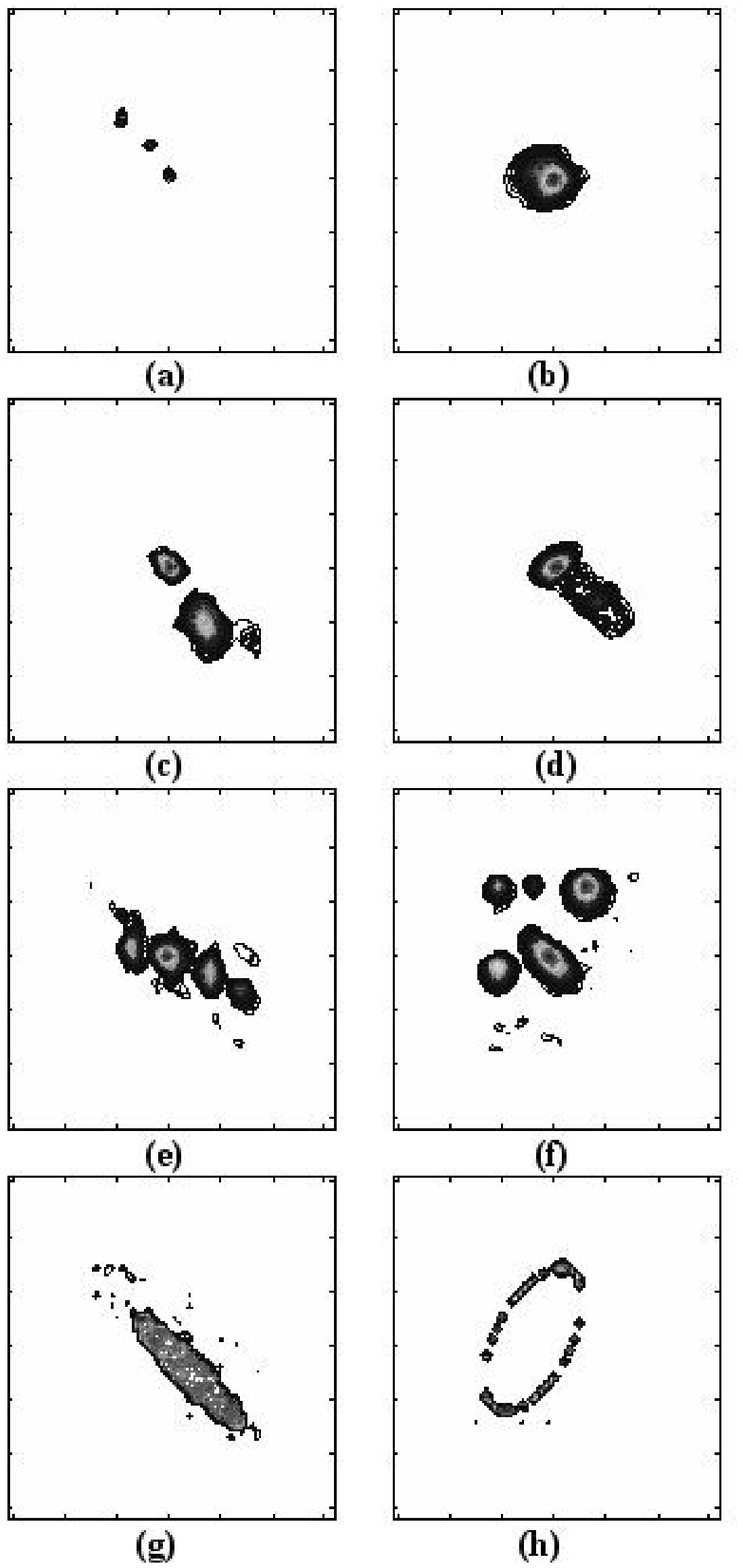,angle=0,width=100mm} }

\vskip 2mm

\begin{center}{{\bf Fig.3.}~Images of the model radio sources constructed \\using
the proposed phase-reconstruction method.}
\end{center}
\end{figure}

\begin{figure}[d]
\centerline{ \psfig{figure=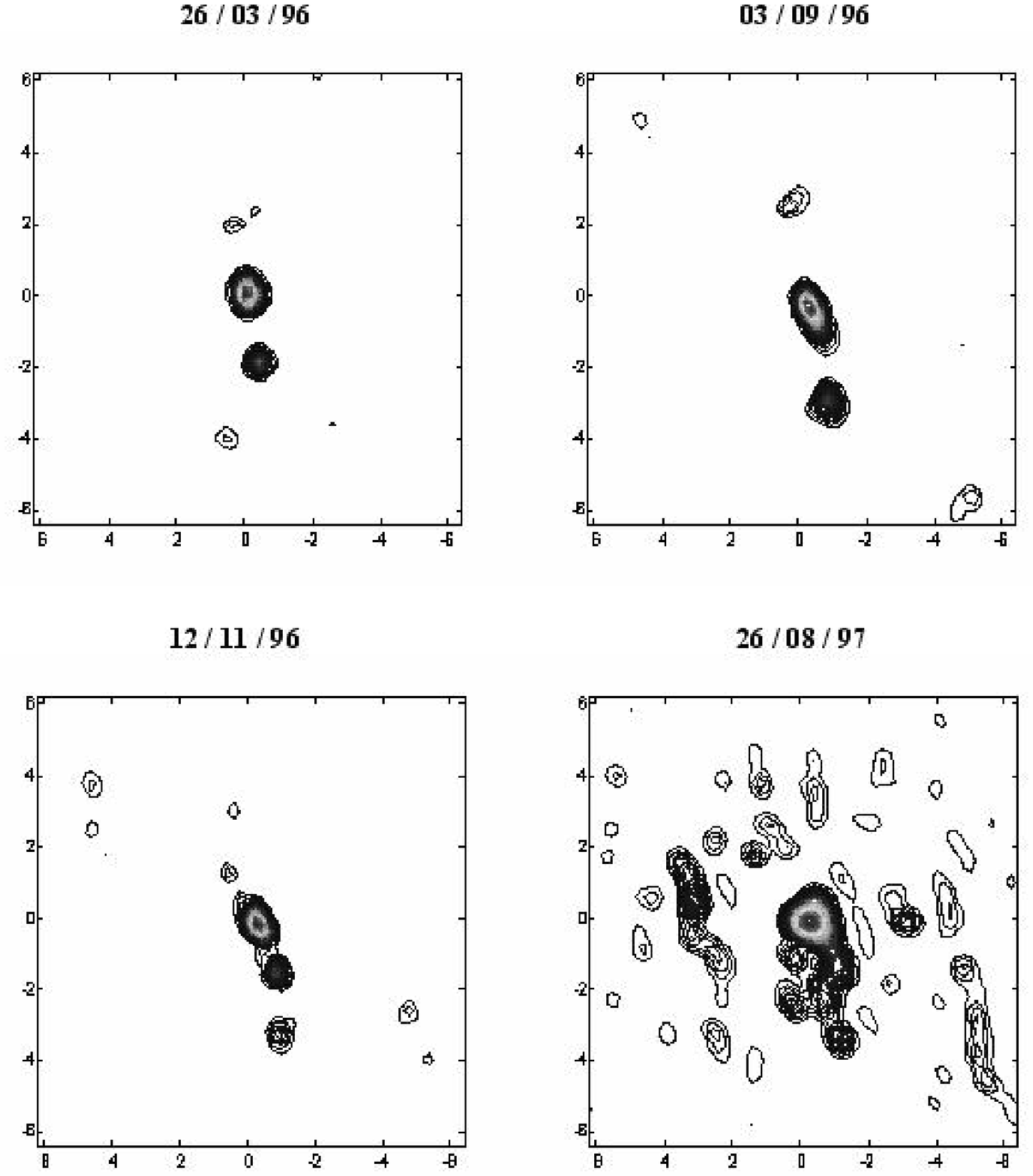,angle=0,width=160mm} }

\vskip 2mm

\begin{center}{{\bf Fig.4.}~Results of mapping the radio source
2200+420\\
using data from a global VLBI array.The scales along the axes are
in mas.}
\end{center}
\end{figure}
\medskip

\begin{figure}[d]
\centerline{ \psfig{figure=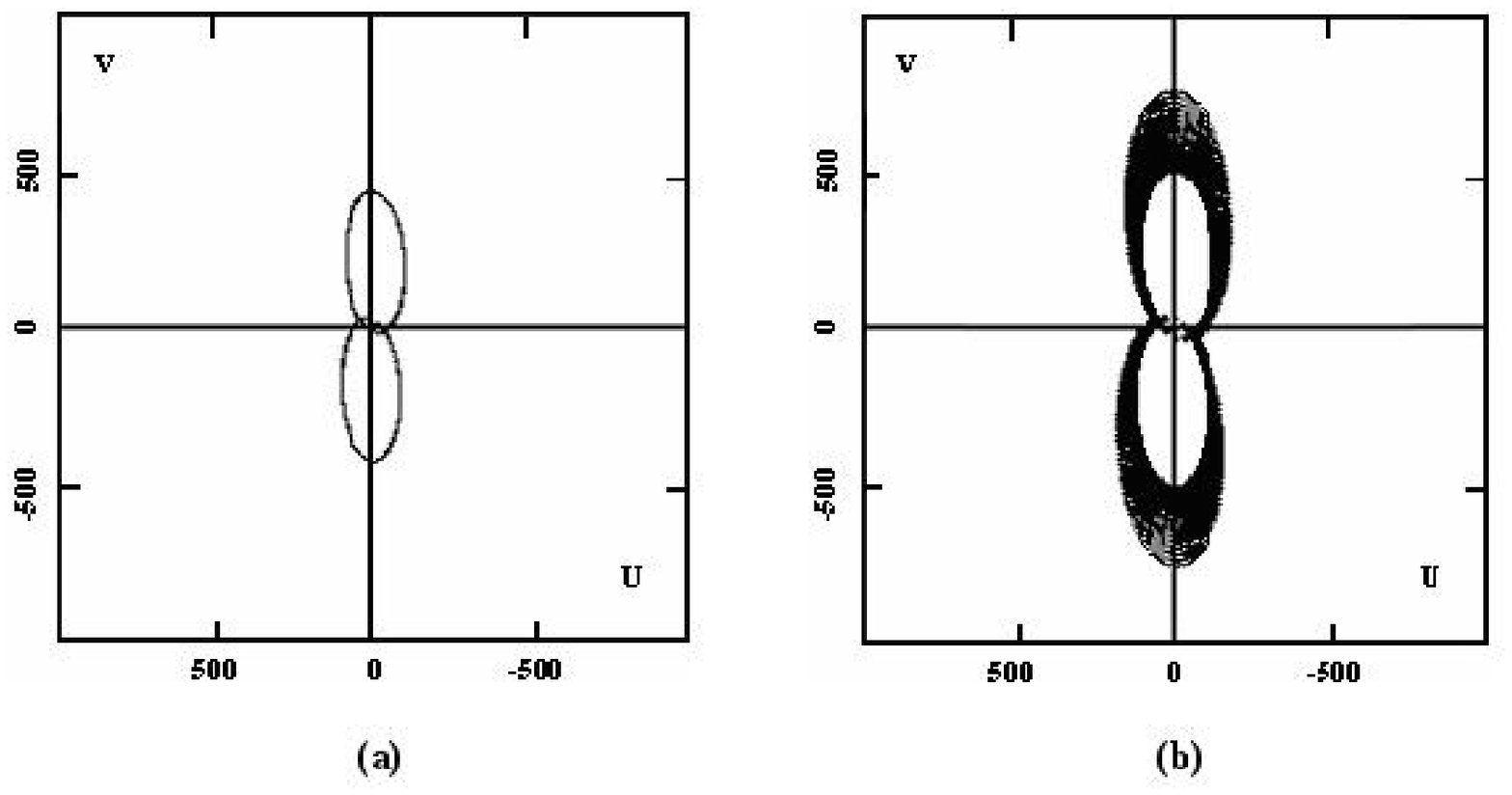,angle=0,width=160mm} }

\vskip 2mm

\begin{center}{{\bf Fig.5.}~Simulations of a ground--space interferometer
for \\(a) single-frequency synthesis and (b) multifrequency
synthesis. \\The scales along the axes are in units of $10^8$
wavelengths.}
\end{center}
\end{figure}
\medskip

\begin{figure}[d]
\centerline{ \psfig{figure=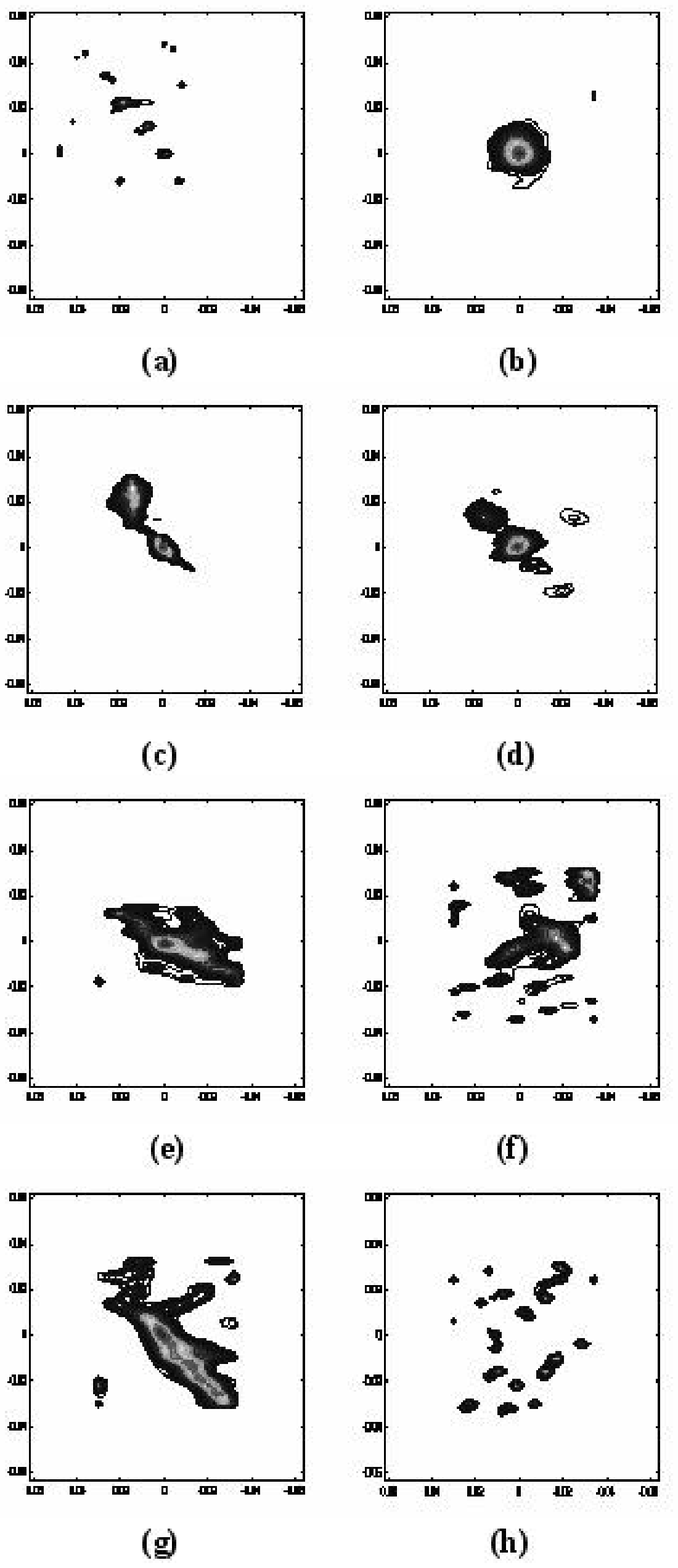,angle=0,width=100mm} }

\vskip 2mm

\begin{center}{{\bf Fig.6.}~Images of model radio sources constructed
from\\
model single-frequency data for a ground-space interferometer
\\with the parameters of RADIOASTRON. The scales along the axes are
in mas.}
\end{center}
\end{figure}
\medskip

\begin{figure}[d]
\centerline{ \psfig{figure=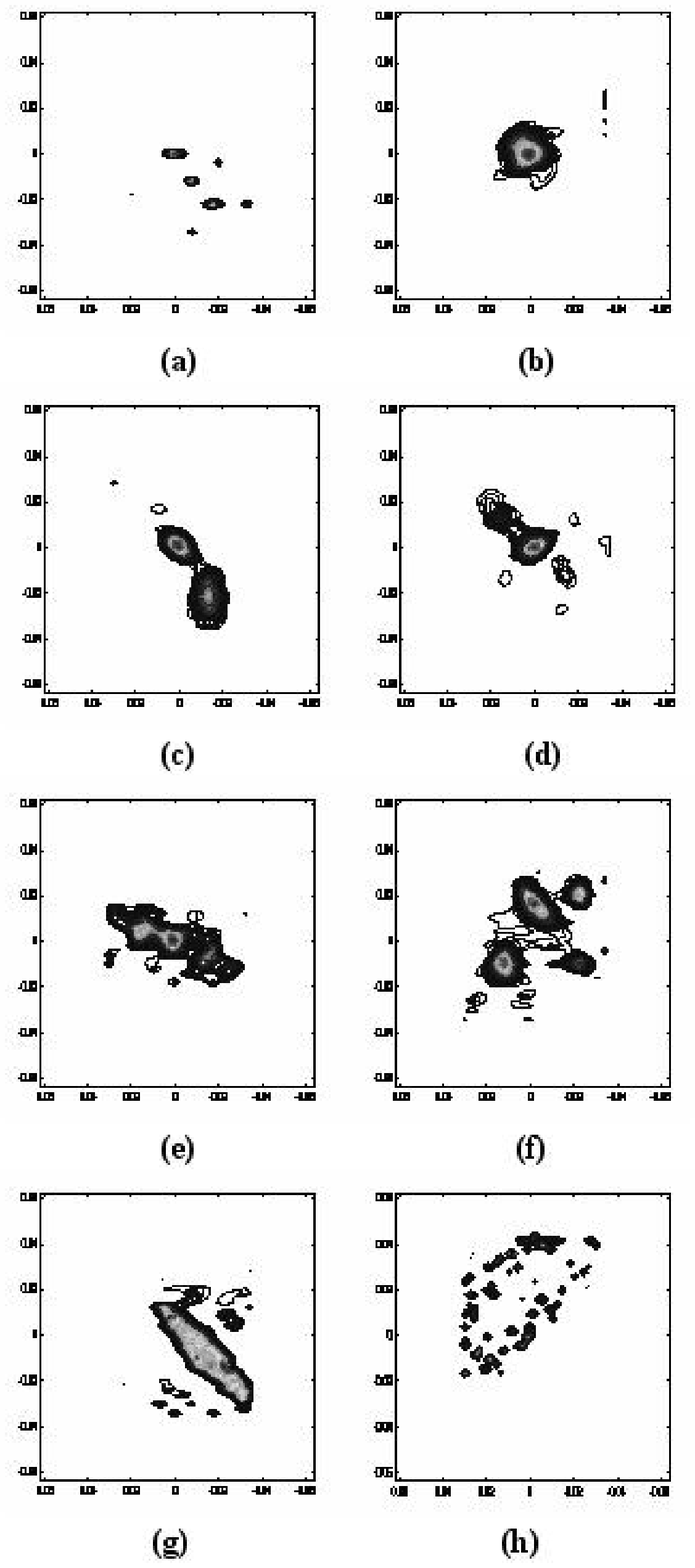,angle=0,width=100mm} }

\vskip 2mm

\begin{center}{{\bf Fig.7.}~Same as Fig.6 but for multifrequency model
data.}
\end{center}
\end{figure}

\end{document}